\documentclass[epj]{svjour}
\usepackage{graphicx}
\usepackage{color}

\newcommand{\ps}{p^\text{s}}

\def\xs{\tilde x(\tau)}

\def\beq{\begin{equation}}
\def\ee{\end{equation}}
\def\bi{\begin {itemize}}
\def\ei{\end{itemize}}

\def\xs{\tilde x(\tau)}

\def\Sm{S_{\rm m}(\tau)}

\def\dst{\dot s^{\rm tot}(t)}

\def\beq{\begin{equation}}
\def\ee{\end{equation}}

\def\bi{\begin {itemize}}
\def\ei{\end{itemize}}

\def\F{{
F}}

\begin{document}

\title{Stochastic thermodynamics of single enzymes and
molecular motors}
\author{Udo Seifert
}

\institute{
{II.} Institut f\"ur Theoretische Physik, Universit\"at Stuttgart,
  70550 Stuttgart, Germany}
\abstract{
For a single enzyme or molecular motor operating
in an aqueous solution of
non-equilibrated
solute concentrations, a thermodynamic description is
developed on the level of an individual trajectory of transitions between
states. The
concept of internal energy, intrinsic entropy and free energy for
states follows from a microscopic description using one assumption on
time-scale separation. A first law energy balance  then allows the
unique identification of the heat dissipated in one transition.
Consistency with the second law on the ensemble level enforces both
stochastic entropy as third contribution to the entropy change
involved in one transition and the local detailed balance condition
for the ratio between forward and backward rates for any transition.
These results follow without assuming
weak coupling between the enzyme and the
solutes, ideal solution behavior or mass action law 
kinetics. The present approach highlights both the crucial role of the intrinsic entropy
of each state and the physically questionable role of chemiostats
for deriving the first law for molecular motors subject to an
external force under realistic conditions.
}


\maketitle

\def\Fe{F^{\rm enz}}
\def\Fn{F_n^{\rm enz}}
\def\Fm{F_m^{\rm enz}}
\def\En{E_n^{\rm enz}}
\def\Em{E_m^{\rm enz}}
\def\Sm{S_m^{\rm enz}}
\def\Sn{S_n^{\rm enz}}

\def\Cn{{\cal C}_n}
\def\k{k_{\rm B}}
\def\kT{\k T}
\def\x{{\bf \xi}}
\def\peqn{p^{\rm eq}_n}
\def\peqm{p^{\rm eq}_m}
\def\peqx{p^{\rm eq}(\xi)}
\def\Ee{{\cal E}^{\rm eq}}
\def\Feq{{\cal F}^{\rm eq}}
\def\Se{{\cal S}^{\rm eq}}
\def\qmn{q_{mn}}
\def\demn{\Delta E_{mn}}
\def\Smn{\Delta S^{\rm}_{mn}}
\def\peqnt{p^{\rm eq}_{n(t)}}
\def\V{V(\x)}

\def\Fs{{F}^{\rm sol}}
\def\Es{{E}^{\rm sol}}
\def\Ss{{S}^{\rm sol}}

\def\mr{{n_\rho^-}}
\def\nr{{n_\rho^+}}
\def\ri{r^\rho_i}
\def\si{s^\rho_i}

\def\mie{\mu_i^{\rm eq}}
\def\deqr{\Delta \mu^{\rm eq}_\rho}

\def\xe{\xi^{\rm enz}}
\def\xs{\xi^{\rm sol}}
\def\xxs{\{\xs\}}
\def\vs{V^{\rm sol}(\xs)}
\def\vx{V(\xe,\xs)}
\def\vt{V^{\rm tot}(\xe,\xs)}
\def\ps{p(\xs)}
\def\pn{p(\xi|n)}
\def\sx{\sum_{{\xs}}}
\def\sn{\sum_{\xi \in {\cal C}_n}}

\def\ftn{F^{\rm }_n}
\def\etn{E^{\rm }_n}
\def\stn{S^{\rm }_n}

\def\ftm{F^{\rm }_m}

\def\fhn{\hat F_n^0}
\def\ehn{\hat E_n^0}
\def\shn{\hat S_n^0}

\def\fhm{\hat F_m^0}

\def\nt{N_T}
\def\nd{N_D}
\def\np{N_P}

\def\Ni{\{N_i\}}
\def\ci{\{c_i\}}

\def\cie{\{c_i^{\rm eq}\}}

\def\rr{_\rho}
\def\dfr{\Delta F\rr}
\def\dfer{\dfr^{\rm enz}}
\def\dfsr{\Delta{F}\rr^{\rm sol}}
\def\der{\Delta E\rr}
\def\deer{\der^{\rm enz}}
\def\desr{\Delta{E}\rr^{\rm sol}}
\def\dsr{\Delta S\rr}
\def\dser{\dsr^{\rm enz}}
\def\dssr{\Delta{ S}\rr^{\rm sol}}

\def\dmr{\Delta \mu_\rho}

\def\ss{^{\rm sys}}

\section{Introduction}
\label{sec1}

Conformational changes of single enzymes have become observable through a variety of
methods often summarized as single molecule techniques \cite{selv07,rito06}. 
Typically, such an enzyme 
 is embedded in an aqueous solution containing different
solutes at specified concentrations. Such a preparation, despite having a well-defined
temperature, often leads to a non-equilibrium system since the aqueous solution
is not in equilibrium with respect to chemical reactions catalyzed by the enzyme.
Moreover, for the case of motor proteins \cite{howard,schl03}, time-dependent
external forces arising  if beads in 
optical traps  are connected via polymeric spacers
 to the enzyme
provide a source of non-equilibrium.

Taking seriously both  the single molecule set-up showing individual transitions
and the  thermodynamic description of the surroundig solution then
prompts the question whether, and if yes how, the thermodynamic laws can be
applied to such processes on the single molecule level. Is it possible to identify
the amount of heat released into (or taken up from) the aqueous solution
constituting an effective heat bath if a molecular motor advances one step? And how
much entropy is produced in such a single step?

The conceptual and technical tools required for such an approach on the level of
individual trajectories have been developed under the label of ``stochastic
energetics''
\cite{seki10}, 
``thermodynamics of
small systems''\cite{bust05},  or, if entropy production is included,
``stochastic thermodynamics'' \cite{seif07},
 recently. The basic idea, indeed, is to
formulate, on the level of an individual trajectory, both a first law, i.e. an energy
conserving balance between an appropriately defined external work, internal energy and
dissipated heat \cite{seki98}, and to identify entropy production
\cite{seif05a}.

The paradigm for such  a  trajectory based approach are colloidal particles in time-dependent
potentials created by various forms of laser traps. Several joint studies between experiments
and theory have illustrated how the thermodynamic quantities can be extracted from records
of the fluctuating trajectories of such driven Brownian motion 
\cite{wang02,carb04,trep04,blic06,gome09,mehl10}.
In a slight generalization, (bio)polymers with their internal shape degrees of freedom
have been subject to a similar analysis some of which directed to deduce free energy
landscapes from  non-equilibrium experiments 
\cite{jarz97,croo99,humm01,liph02,coll05,impa07b,rito08,juni09}. 
In both cases, the theoretical description 
typically is 
based on Langevin-type dynamics.

Molecular motors have also been modelled using 
Lange- vin dynamics in a
potential that depends explicitly on the current chemical state of the motor
\cite{juel97,astu02,reim02a,parr02}. Alternatively, as often used for enzymes,
a description based on discrete, distinguishable states between which (sudden) transitions
take place is often more appropriate 
\cite{fish99,lipo00,bust01,bake04,andr06c,gasp07,liep07,liep07a,liep08,liep09,lipo09,lau07a,kolo07,astu10}. 
In most of these works the focus has been on elucidating the
cycles involved in the action of the motor and on deriving force-velocity curves and their
dependence on ATP and ADP concentrations. A stochastic thermodynamics approach
has been applied in Refs.
\cite{andr06c,lau07a} for deriving a fluctuation theorem and in 
Refs. 
\cite{liep07,liep07a,liep08,liep09,lipo09} for the identification of heat  
 on the trajectory level. Thermodynamics along a trajectory was discussed for 
a
 single
enzyme
in Refs. \cite{seif05,schm06a,min05} and for 
chemical reaction networks in Refs. \cite{shib00,gasp04,schm06}.

The purpose of the present paper is to develop the stochastic thermodynamics of single enzymes
and molecular motors afresh and coherently
under minimal assumptions thus clearly dissecting conditions that are 
necessary from those which are (too) simplifying but not necessary. Since this work deals 
partially with topics addressed previously it is appropriate to point out the main differences
and new results up front. 
First, some of the earlier work (and even recent ones \cite{ge10}) does not distinguish
carefully between internal energy and free energy of a state. Enzymes differ in this respect 
from simple colloidal
particles which have no relevant internal degrees of freedom. 
Within stochastic thermodynamics the first recognition of this aspect
seems to have been our discussion of the role of degenerate states \cite{schm06a}
 but a more systematic approach
starting from a solid microscopic model is appropriate.  A consequence of the correct 
treatment is that the heat released in one step can no longer be inferred from the rates 
directly as it can in the case of a colloidal particle. 
Moreover, formulated correctly, heat in the stochastic thermodynamics approach
becomes 
unique and identical to
the caloric one
thereby remeding a deficiency pointed out by Sekimoto \cite{seki10,seki07}. 
Second, the thermodynamics of molecular motors has previously been described 
using the concept of ``chemiostats''
\cite{liep07,liep07a,liep08,liep09,lipo09}.
 We show that the heat thus identified would
 require
rather unrealistic experimental conditions and should, in practice, be replaced by a new expression
derived below.
Third, in our earlier work \cite{schm06a,schm06}, explicit expressions for the 
individual transition rates based on mass action law kinetics
have been invoked early on thus effectively restricting the range of applicability
unnecessarily. In fact, the first law can be formulated on the trajectory
level without any assumptions on the transition rates. Finally, we now can show that 
the correct identification of entropy production occuring in one step follows 
quite naturally from requiring consistency with the second law on the ensemble level. 
In particular, this 
condition leads to both stochastic entropy \cite{seif05a} as a necessary
third contribution to the entropy
change associated with one transition  and
to an expression for the ratio of the forward and backward rate known as local detailed 
balance. The latter property is usually either postulated or claimed to follow from
microreversibility which is a concept  not so trivially applicable to chemical 
transformations under non-equilibrium conditions.

This paper is organized as follows. In Sect. \ref{sec2}, to set the stage, we define thermodynamic 
notions for the states of an enzyme undergoing only conformational changes and not yet
catalyzing reactions. In Sect. \ref{sec3}, we discuss the modifications required if we allow
chemical reactions. In Sect. \ref{sec4}, we apply this formalism to molecular motors. 
 Sect. \ref{sec5} deals with entropy production for all these systems. A brief summary and conclusions
follows in Sect. \ref{sec6}. In the appendix, we 
discuss the implications of our trajectory-based approach for the thermodynamics of 
time-dependent ensembles.

\section {Configurational transitions of a single enzyme}
\label{sec2}
\subsection{Solution}
The enzyme will be placed in an aqueous solution which consists of a set of
$\{N_i\}$
molecules of type $i$ enclosed in a volume $V$ at a  temperature
$T$. 
The microscopic configurations, i.e., the micro-states, of this 
solution (without the enzyme yet)
 are labelled collectively by $\xxs$. The
configurational energy of the whole solution can be expressed by a 
potential $\vs$ leading to the probability 
\beq
\ps = \exp[-\beta(\vs+\Fs)]
\ee
for each micro state $\xs$ with the free energy
\beq
\Fs\equiv -\k T\ln\sx  \exp[-\beta\vs] .
\ee 
Here $\beta \equiv 1/\k T$ is the inverse temperature and $\k$ Boltzmann's constant.
The (mean) internal energy
of this solution is given by
\beq
\Es=\sx\ps\vs
\ee
and its entropy by
\beq
\Ss\equiv -\k\sx\ps\ln\ps =(\Es-\Fs)/T  .
\ee
All these quantities depend on $T,V$ and $\{N_i\}$.
Moreover, we will
assume that this solution is  large enough to be treated
in the thermodynamic limit which implies that  
the chemical potential for species $i$, 
\beq
\mu_i =\partial_{{N_i}}\Fs,
\label{eq:mu}
\ee
becomes a function of $T$ and
the concentrations $\{c_i\}\equiv \{N_i/V\}$ only. 

\subsection{Thermodynamic quantities  of states}

\begin{figure}
\includegraphics{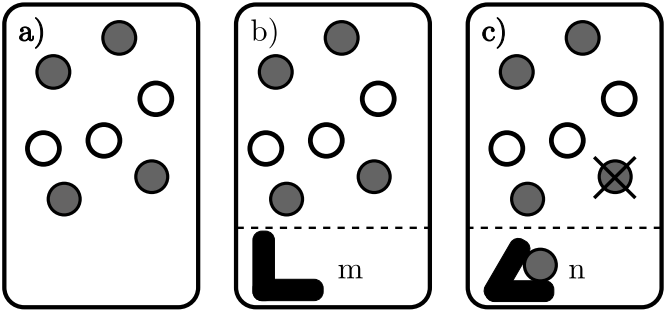}
\caption{(a) Solution with two types of solutes. (b) Enzyme in solution 
in state $m$. (c) Enzyme in state $n$ with one solute molecule tightly bound,
and, consequently one less molecule in solution.
The dashed line in (b) and (c) reflects the
purely formal splitting of the total system into the solution and the enzyme
as indicated in (\ref{eq:fn}) and (\ref{eq:fn-bind}) 
for the free energy of state $m$ and state 
$n$, respectively.}
\end{figure}

To this solution, we add 
a single enzyme, see Fig. 1.
Following in spirit Hill's classical work
\cite{hill}, 
we distinguish different (mesoscopic) states of the enzyme 
 such that equilibration
among microstates corresponding to the same  state
is fast whereas transitions between these states are assumed to be
slower and observable. Under these conditions, we can assign to
each state $n$ a free energy $\Fn$, an internal energy $\En$, and
an intrinsic entropy $\Sn$ following in spirit Hill's classical work
\cite{hill}.

We denote the microscopic
configurational degrees of freedom of an enzyme with fixed position
of its center of mass collectively by $\{\xe$\}.
The full  configurational energy of the system consisting of 
enzyme and solution becomes
\beq
\vt\equiv \vs +\vx \equiv V^{\rm tot}(\xi) ,
 \ee
where $\vx$ contains both the interaction within the enzyme and
the interaction between enzyme and solution. 
 We now
partition all microstates $\{\xi\}=\{(\xe,\xs)\}$ of the combined system
enzyme and solution
 into a set of state configurations
$\{\Cn\}$ such that each microstate $\xi$  of the combined system
occurs in one and only one
such set $\Cn$. For any specific state $n$, the probability $\pn$ of
finding an allowed microstate of the combined system consisting of
enzyme and solution then follows from the assumption
of fast equilibration as
\beq
\pn= \exp[-\beta(V^{\rm tot}(\xi) -\ftn)] 
\ee
with $\beta\equiv 1/\kT $ and the constrained  free energy in 
 state $n$
\beq
\ftn \equiv -\kT \ln \sn \exp[-\beta V^{\rm tot}(\xi)]
\label{eq:fnn}
\ee
ensuring proper normalization $\sn\pn=1$.
The    (mean) internal energy in state $n$ is
\beq
\etn\equiv   \sn \pn V^{\rm tot}(\xi)
\ee and
the (intrinsic) entropy becomes as usually
\beq
\stn = -\k \sn\pn\ln \pn=(\etn-\ftn)/T 
\label{eq:snn}
 .
\ee

The thus defined free energy, internal energy and (intrinsic)
entropy of each 
state of the combined system will depend on $T,V$ and $\{N_i\}$.  For a finite range of the interaction
potential 
$\vx$, we can indentify the
free energy, internal energy and intrinsic entropy of the enzyme proper
 with
\beq
\Fn(\ci)\equiv \ftn(\Ni)-\Fs(\Ni) ,
\label{eq:fn}
\ee
\beq
\En(\ci)\equiv \etn(\Ni)-\Es(\Ni) ,
\ee
and
\beq
\Sn(\ci)\equiv \stn(\Ni)-\Ss(\Ni) 
\label{eq:sn} ,
\ee
respectively. In the thermodynamic limit of the
solution, these quantities become independent of system size and  depend only on the
concentrations $\{c_i\}$. Since we keep $T$ and $V$ fixed throughout the paper,
we suppress the dependence on these quantities notationally and
often that on $(\{c_i\})$ as well.

Since both the full  quantities as well as the bare solution
quantities obey the usual relation between free energy, internal energy and
entropy, it is obvious that
\beq
\Fn=\En-T\Sn 
\label{eq:fes}
\ee
holds as well.
 So it looks like the full system, enzyme plus solution,
could be split into  two subsystems with additive thermodynamic quantities
despite the fact that the enzyme is neither a macroscopic object nor that we
have assumed that the interaction between enzyme and solution is in
any sense small.
The caveat, however, is that
the quantities refering to the enzyme ($\Fn,\En$ and $\Sn$) will depend on the concentrations
$\{c_i\}$ which refer primarily to  properties of the solution.

\subsection{First law}

In the spirit of stochastic thermodynamics, we now formulate
a first-law like energy balance for transitions between states.
 Obviously,
there is no external work playing any role for such a closed system.
If the enzyme jumps from state $m$ to state $n$, the change in 
internal energy
\beq
\Delta E \equiv E_n-E_m =\En-\Em= -q
\label{eq:qe}
\ee
can be identified with an amount $q$ of heat being released into
(or, if
negative, being taken up from) the surrounding heat bath.

\section{Enzymatic reactions}
\label{sec3}

\subsection{Binding of substrate molecules}

For an enzyme, configurational changes often involve the binding
or release of smaller molecules from the surrounding solution
like the nucleotides ATP, ADP or inorganic phosphate P$_
{\rm i}$.
For  states with such bound molecules, like the one shown in Fig. 1c, 
 it will be convenient to 
identify  the free energy of  the state somewhat differently than
done in  (\ref{eq:fn}).
In the case where  state
$n$ has one $A_1$ molecule tightly bound to it, we 
define
\beq
\Fn\equiv  \ftn(N_1) -\Fs(N_1-1) 
=\ftn(N_1)+ \mu_1 - \Fs(N_1) ,
\label{eq:fn-bind}
\ee
where we use (\ref{eq:mu}) and 
drop notationally the dependence on the irrelevant species 
$\Ni$ with $i\not =1$.
This definition thus means that $\Fn$ is obtained by subtracting from the
total free energy $\ftn$ of the combined system
the free energy of a solution containing one
less $A_1$  molecule (the bound one). The idea behind it
is  again a conceptual (but not physical) 
splitting of the whole system into the enzyme (plus the bound molecule)
and the solution
 possible despite the fact that both may interact strongly.

The advantage of this scheme becomes obvious if we consider a binding
event conventionally written as
\beq
m + A_1 \rightleftharpoons n
\ee
where upon binding of an $A_1$ molecule a state $m$ transforms into
the  state $n$ just discussed, compare Fig. 1 (b) and (c). 
The free energy difference involved in this process
becomes
\begin{eqnarray}
\Delta F &=& \ftn -\ftm \\&=& (\Fn-\mu_1+ \Fs) - (\Fm+\Fs) \\
&=& \Fn-\Fm-\mu_1 .
\label{eq:DelF}
\end{eqnarray}
If one changes the concentrations of $A_1$ molecules in the solution
one would expect that the difference in free energies depends on this concentration.
Such a concentrations dependence becomes obvious through the $\mu_1$ 
term. The free energy  $\Fn$ defined according to (\ref{eq:fn-bind})
 will typically only weakly depend on concentration since it
mainly
contains the interaction between the one bound $A_1$ and the enzyme.
If we had used the definition (\ref{eq:fn}) also for the free
energy of the enzyme in   state $n$,
$\mu_1$ would not show up explicitly in (\ref{eq:DelF}). 
The dominant concentration dependence
of the free energy change would then be hidden in the $\Fn$ term.

In a more general case, if a  state $n$  has $\sum_i r_iA_i$
molecules bound to it, we define its free energy as
\begin{eqnarray}
\Fn(\ci)&\equiv&  \ftn(\{N_i\}) -\Fs(\{N_i-r_i\})
   \label{eq:Ftwo}  \\
&=&\ftn(\{N_i\})+ \sum_i r_i\mu_i - \Fs(\{N_i\}) ,
\end{eqnarray}
Likewise, one can identify the  entropy and internal energy
of the enzyme in such a state  accordingly as
\begin{eqnarray}
\Sn(\ci)&\equiv& 
\stn(\{N_i\}) -\Ss(\{N_i-r_i\}) \\
&=&\stn(\{N_i\})- \sum_i r_i\partial_T\mu_i - \Ss(\{N_i\}),
\end{eqnarray}
where we use 
\beq
\partial_{N_i}\Ss=-\partial_T\mu_i,
\ee
and
\begin{eqnarray}
\En(\ci)&\equiv&  \etn(\{N_i\}) -\Es(\{N_i-r_i\})\\
&=&\etn(\{N_i\})+ \\&~&~~+ \sum_i r_i (\mu_i-T\partial_T\mu_i) - \Es(\{N_i\})
\label{eq:Stwo}
\end{eqnarray}

So from now on for the precise definition of the free energy, internal
energy and entropy  of a  state
we need to distinguish between states that have solute molecules bound
to them for which the definition   (\ref{eq:Ftwo}-\ref{eq:Stwo}) will be used from those 
which have not for which we will use (\ref{eq:fn}-\ref{eq:sn}). 
While this distinction of identifying the quantities with the
superscript ``${\rm enz}$'' may seem somewhat pedantic 
for pure binding reactions, it becomes mandatory when
we consider enzymatic reactions.

\subsection{Example: Hydrolysis of ATP}

\def\at{{\rm ATP}}
\def\ad{{\rm ADP}}
\def\ap{{\rm P_{\rm i}}}

 For a typical example involving an enzymatic reaction
 consider binding
and subsequent hydrolysis of ATP in a solution containing
 ATP, ADP and P$_{\rm i}$, at chemical potentials $\mu_T, \mu_D,$ and
$\mu_P$,
respectively. The enzyme undergoes a transition
\beq
\underbrace {k + \at}_{1} \rightleftharpoons \underbrace {(m,\at)}_{2} 
\rightleftharpoons  \underbrace {(n,\ad,\ap)}_{3} 
\rightleftharpoons  \underbrace { k + \ad + \ap}_{4} .
\label{hydro}
\ee 
We have made explicit that in  state $m$ an ATP and that in 
state $n$ an ADP and a $\ap$ are tightly bound to the enzyme.
The overall
 reaction becomes
\beq
 k + \at \rightleftharpoons k +  \ad + \ap .
\ee
For simplicity, we have constrained the enzyme to be in the same state
$k$ (without bound molecules) before and after the reaction.
The free energy difference for the first step involving the
 binding of the
ATP becomes
\beq
F_2-F_1=(\Fm-\mu_T)-F_k^{\rm enz},
\ee
where we use  (\ref{eq:Ftwo}) since state $m$ has an ATP bound to it.
Likewise, the free energy difference upon release of ADP and $\ap$
becomes
\beq
F_4-F_3=F_k^{\rm enz}-(\Fn-\mu_D-\mu_P).
\ee
Since the overall free energy difference clearly is
\beq
\Delta F =F_4-F_1 = \mu_D + \mu_P -\mu_T ,
\ee
we obtain for the free energy difference between the two
intermediate states the expression
\beq
F_3-F_2=(F_3-F_4) + ( F_4-F_1) + ( F_1-F_2)=  \Fn -\Fm.
\ee
The last line shows the advantage of expressing the free energy
of the enzyme in terms of definitions (\ref{eq:Ftwo}). Clearly, 
the free energy difference between these two intermediate states 
should not strongly depend on concentrations of
the solutes, i.e., should not
contain terms
that depend explicitly
on their chemical potentials.

\subsection {General case}

For a general case, consider  transitions typically
written as
\beq
\mr + \sum_i \ri A_i  \rightleftharpoons 
\nr + \sum_i \si A_i
\label{reactm}
\ee
where $1\leq \rho \leq N_\rho$ labels the possible transitions.
Here, $\mr$ and $\nr$ denote the states of the enzyme
before and after the reaction, respectively.
Following the scheme just applied to the example above, we obtain
for the free energy difference 
involved in this transition 
\beq \dfr\equiv\dfer +\dfsr
\ee
where 
\beq
\dfer\equiv \Fe_\nr-\Fe_\mr
\ee
denotes the free energy change of the enzyme and
\beq
\dfsr = \sum_i(\si-\ri)\mu_i\equiv \Delta \mu_\rho
\label{eq:dmr}
\ee
denotes the free energy change attributed to the solution in this reaction.
Note that these relations remain true even if the
 states $\mr$ and $\nr$ have both  the same additional
molecules not showing up in (\ref{reactm}) 
 bound to them provided one then uses for their free energies
the definition analogously to   (\ref{eq:Ftwo}).
Likewise, the change in internal energy and entropy of the combined system becomes
\begin{eqnarray}
\der&\equiv& \deer+\desr 
\label{eq:der}
\\
&=& E_{\nr}^{\rm enz}-E_{\mr}^{\rm enz} +\dmr-T\partial_T\dmr, 
\end{eqnarray}
and
\begin{eqnarray}
\dsr&\equiv& \dser+\dssr\label{eq:dsr}
\\
&=&S_{\nr}^{\rm enz}-S_{\mr}^{\rm enz}  -\partial_T\dmr ,
\end{eqnarray}
respectively.

\subsection{First law}

As in the case of pure conformational changes, we now want to assign a
first law type energy balance to each reaction of type $\rho$ shown in 
({\ref{reactm}). Once an initial
state
is prepared, in the closed system (enzyme plus solution) there is obviously
no source of external work. Neither does the system perform  any work.
Hence, the heat released in this transition is given by minus the change of internal
energy of the combined system (\ref{eq:der})
\beq
q\rr =  -\der =-\deer
-\dmr + T \partial_T \dmr. 
\label{eq:heatrelax}
\ee
This relation shows  that the enzyme and the solution are treated on the
same footing since only their combined  change in internal energy
enters. Clearly, since
the heat is released into the solution acting as a thermal bath, the 
configurational change of the enzyme
as well as  binding and releasing solute molecules contribute to the same bath.

\section{Molecular motors}
\label{sec4}
\subsection{First law}
\def\wm{w^{{\rm mech}}_{\rho}}

Essentially the same formalism
applies to an enzyme acting as a molecular motor often described by such
discrete states. 
Most generally, if the motor undergoes a forward 
transition of type $\rho$ 
as in
(\ref{reactm}) it may advance a distance $d_\rho$ in the direction of
the applied force $f$ (or, if $f<0$, opposite to it). We allow the special cases $d_\rho =0$ (pure
chemical step) or  $\si=\ri=0$ (pure mechanical step) but do
not exclude that both types are involved in one transition.
 For $d_\rho \not = 0$, the 
mechanical work 
\beq
\wm\equiv fd\rr
\ee
is  applied to (or, if negative, delivered by)  the
motor.
 
We first consider the case that the motor is operating in  an
environment where the concentration of molecules
like ATP, ADP or P$_{\rm i}$ are initially fixed. Effectively, these conditions
correspond to a closed system as discussed above for an enzymatic reaction. 
In an almost trivial
extension of (\ref{eq:heatrelax}) the first
law for a single transition of type $\rho$ becomes
\beq
q_\rho = \wm - \Delta E_\rho 
= fd_\rho -\deer -\dmr +T\partial_T\dmr  .
\label{eq:qmot}
\ee

\subsection{Comparison to previous work: ``Chemiostats''}

The form (\ref{eq:qmot}) of the first law with the concomitant identification of the
heat dissipated in such a transition is original to the present work. It
differs from the form discussed previously for molecular motors by Baker
\cite{bake04}, 
and, more recently,
in particular by 
Lipowsky and co-workers \cite{liep07,liep07a,liep08,liep09,lipo09}. 
In their work, the first law for a step like in (\ref{reactm})
is formulated (using our notation and sign convention) as
\beq
\bar q_\rho = \wm -\deer -\dmr
\ee
where we use the overbar to distinguish their heat
\beq
\bar q_\rho = q_\rho - T\partial_T \dmr = q_\rho + T \dssr
\ee from the present $q_\rho$. If the heat released in one step is a physically meaningful concept, it should
be
unique. Hence, only one expression, either $q_\rho$ or $\bar q_\rho$,
 can be the correct one.

Formally,
the two expressions for the heat differ by a term involving the entropy change in the
solution resulting from the reaction. 
The physical origin of the two different forms
arises from the fact that in the previous work
the enzyme is thought to be coupled to ``chemiostats'' providing and
accepting molecules at an energetic cost (or benefit) given by
their chemical potential. Introducing
the notion of a chemical work 
\beq
w_\rho^{\rm chem} \equiv  - \Delta \mu_\rho
\ee
the first law is then written in the form
\beq
\wm + w_\rho^{\rm chem} = \deer + \bar q_\rho  .
\ee

The origin of the difference between the two approaches becomes  clear
 by analyzing the operation of chemiostats in more detail in the
context of whether the
concentration of the $A_i$ molecules are kept
 strictly constant or not. This 
distinction has been alluded to in our previous work on enzymatic
reactions \cite{schm06a} and biochemical recation networks \cite{schm06}
 where the same subtlety arises but it
seems appropriate to provide a more explicit and detailed discussion
in order to settle this important point. Essentially, one has to
distinguish two different scenarios.

Scenario I is the one discussed in the present paper so far where we prepare a
non-equilibrium state by selecting non-equilibrium concentrations $\{c_i\}$ and
then let
the motor run. The first law in the form (\ref{eq:qmot}) and the corresponding
identification
of the heat then seems inevitable. A side effect of such a set-up, however, is
the
fact that strictly speaking the concentrations $\{c_i\}$ will (slowly) change in a 
finite system. Insisting on a strictly constant concentration will lead
us to scenario II.

In this second scenario,  
one wants to control
the concentrations of these solute molecules throughout the experiment.
Literally speaking, one then has to refill or extract certain
molecules after a reaction event has taken place. Practically, this
can obviously not be done in any strict manner. Conceptually, however,
we can conceive devices, which are effectively the chemiostats,  that ``reset'' the number of solute
molecules after each step. Of course, such an intervention has to
obey a first law as well which we formulate for such a reset
operation following the reaction of type $\rho$ as
\beq
w^{\rm re}_\rho  =  \Delta {E}^{\rm sol,re}_\rho     + q^{\rm re}_\rho
=-\desr +   q^{\rm re}_\rho,
\label{firstreset}\ee
where the superscript ``${\rm re}$'' stands for reset. The change in internal energy of the solution
 $\Delta {E}^{\rm sol,re}_\rho$ is minus
the corresponding change in internal energy of the solution, i.e., 
$-\desr$,
  when the
reaction took place.
If this reset operation occurs quasistatically, the work $w_\rho^{\rm re}$
spent
in it is equal to the free energy change of the solution in this operation,
which is $-\Delta \mu_\rho $, 
leading to the identification
\beq
 q^{\rm re}_\rho = -\dmr + \desr= T\dssr = \bar q_\rho  - q_\rho. 
\label{heatreset}
\ee

Hence, the heat $\bar q_\rho$ discussed in previous works  for a single step under chemiostatted
conditions
physically would correspond to the sum of (i) the heat  $q_\rho$ dissipated in the
reaction step and (ii) the heat $q^{\rm re}_\rho$
dissipated in the subsequent
quasistatic steps when the
molecules involved in the reaction are fed in
and taken out. Note that such a procedure makes
sure that the
concentrations  literally have not  changed at all.

While this scenario II may be possible conceptually it is difficult to envisage
a practical experimental implementation. It thus seems that for identifying the
heat the approach taken
in the
present paper is the physically more realistic and relevant one since the
motor accepts and releases molecules directly from the
surrounding solution with no further feed-back-type interference  from chemiostats.

\section{Entropy production}
\label{sec5}
\subsection{Motivation}

Naively, one might have expected that with the correct identification of
both the heat, which should be attributed to a change of the entropy of the
surrounding heat bath, or "medium", via
\beq
\Delta S^{\rm med}_\rho \equiv  q_\rho/T  , 
\label{eq:sm}
\ee
 and the intrinsic entropy change of the system  $\dsr$ as given by
(\ref{eq:dsr}),
the total entropy change in one step is given by the sum of both, i.e., 
$\Delta S  = \Delta S^{\rm med}_\rho + \Delta S $.

For a counterexample showing that such a view would be  too simplistic
consider an enzyme with just two states, $m$ and $n$, with $E_m^{\rm enz}=E_n^{\rm enz}$
and $S_m^{\rm enz}> S_n^{\rm enz}$. If the enzyme is initially in state
$m$ it will at some time jump to state $n$. Such an isoenergetic transition
involves no exchanged heat and hence no change in the entropy of the medium.
Clearly, then the sum of the changes in intrinsic system entropy and medium entropy is
negative for such a transition. While this is not a problem for a single
enzyme it becomes one if we consider a whole ensemble of enzymes
all prepared initially in state $m$. Likewise, we could repeat the
experiment with a single enzyme initially prepared in state $m$
 many times for obtaining an ensemble 
average. In both cases,  naively averaging the total entropy change as tentatively
identified above, we would get on average a decrease in total entropy.
Such a conclusion violates the second law and hence something is missing.
We now show that we have to add a third contribution to the entropy, called
stochastic entropy \cite{seif05a}, in order to achieve a consistent description of
entropy changes  on the level of transitions along
an individual trajectory. This concept necessarily requires an ensemble description from which
the individual trajectories are taken.

On the ensemble level, entropy production in (bio)che- mical reaction 
networks has been investigated for quite some time using master equation
\cite{hill,schn76,luo84,mou86,qian05}. The main point in the following will  not
be to
repeat this analysis but rather to use consistency with the second law
on the ensemble level together with the insight into the first law derived
above for a complete identification of the entropy production to be associated
with an individual transition on the trajectory level.

\def\wrp{w_\rho^+}
\def\wrm{w_\rho^-}
\def\wrpm{w_\rho^\pm}
\def\dsrt{\Delta S^{\rm tot}_\rho}
\def\dst{\Delta S^{\rm tot}}
\def\dsmr{\Delta S^{\rm med}_\rho}
\def\prp{p_\nr}
\def\prm{p_\mr}

\subsection{Ensemble}
In the course of time,  the enzyme will jump between different states.
The jump times will be stochastic since there is only a certain 
probability that a reaction of type (\ref{reactm}) takes place if the enzyme is
in the state $\mr$ corresponding to the left hand side of (\ref{reactm}). 
A trajectory of the enzyme can then be characterized by the sequence of
jump times
$\{t_j\}$ and the sequence of reactions $\{\rho_j^{\sigma_j}\}$ where $\rho_j$
denotes the corresponding reactions and $\sigma_j= \pm$ characterizes the
direction in which the reaction takes place, see Fig. 2 for an example based on
the scheme (\ref{hydro}).

An ensemble
is defined by specifying (i) the initial probability $p_n(0)$ for finding
the enzyme in state $n$ and (ii) the set of rates $\wrpm$ 
with which the reactions (\ref{reactm})
takes place in either direction. Both inputs will then determine the
probability
$p_n(t)$ to find the enzyme in state $n$ at time $t$. 
Averages with respect to such an ensemble will be denoted by
$\langle ...\rangle$.

\begin{figure}
\includegraphics{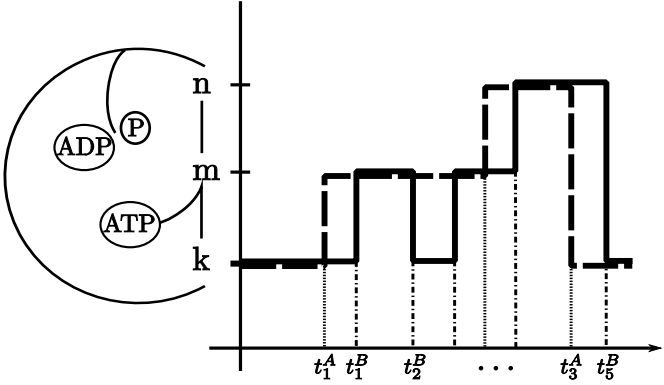}
\caption{Two trajectories A (dashed line) and B (full line) for an enzyme undergoing
the scheme (\ref{hydro}) with the reactions 
(1) $k + ATP \rightleftharpoons m$,
(2) $m  \rightleftharpoons n$, and 
(3) $n \rightleftharpoons k + ADP + P_i$.
For trajectory A, the three transitions are all in forward direction
with
$\rho_j=j$ and $\sigma_j=+1$ for $j=1,2,3$ at jump times $t_j^A$.
Trajectory B exhibits five transitions at times $t_j^B$ with 
$\rho_1=\rho_2=\rho_3=1, \rho_4=2$ and
$\rho_5=3$ with one backward step $(\sigma_2=-1$) and all other $\sigma_j = +1$.
}
\end{figure}

\subsection{Entropy production in one step}

As explained above, naively adding the entropy change of the medium and
the intrinsic one of the system will not necessarily lead to an on
average non-negative entropy production. Therefore, we tentatively
write the total entropy change occuring in a forward transition of type (\ref{reactm})
as
\beq
\dsrt(t) = \dsmr + \dsr + \Delta s_\rho(t)
\label{eq:sss}
\ee
where the last term is the one still to be determined. The reason for introducing an explicite
time-dependence will become clear below. The corresponding
value of the total entropy change involved in  a backward transition would be $-\dsrt(t)$.
As an essential
requirement, we impose the condition that the average total entropy production rate
is non-negative, i.e. that \beq
0\leq\langle \dot S^{\rm tot}(t)\rangle = \sum_\rho [p_\mr(t)\wrp - p_\nr(t)\wrm] 
\dsrt(t) 
\label{eq:ssss}
\ee 
where here and in the following the dot denotes a time-derivative, i.e., a rate.
The explicit expression for this average arises from exploiting the fact
that with probability $p_\mr(t)$ 
the enzyme is in state $\mr$ allowing at time $t$ the reaction $\rho$ to take place 
in forward direction with rate
$\wrp$. Likewise, the enzyme is with probability $p_\nr(t)$
in the state $\nr$
allowing the reaction to proceed in backward direction.

Now suppose that we knew the rates $\wrpm$ and were looking for $\dsrt(t)$
as a function of $p_\mr(t), p_\nr(t)$ and these rates. 
Since the inequality (\ref{eq:ssss})
has to be respected
for any $p_n(t)$, it looks inevitable that each individual term in this
sum has to be  non-negative, i.e., 
\beq
0\leq [p_\mr\wrp - p_\nr\wrm] \dsrt \equiv (y-x) \dsrt(x,y) 
\ee
which defines the abbreviations $x$ and $y$ and 
where we suppress the $t$-dependence notationally.
The yet unknown function $\dsrt$ has dimension of entropy, i.e., we can write
\beq
 \dsrt = k_B f(x,y)
\ee
Since the function $f(x,y)$ is dimensionless,  it can  depend only
on a dimensionless variable, i.e. $f(x,y) = g(y/x)=g(z)$. 
Finally, the requirement that interchanging forward and backward 
directions of the reaction $\rho$, which amounts to interchanging $x$ and $y$,
 corresponds to a sign change in the
entropy  imposes the condition
\beq
g(z) = - g(1/z).
\ee
Up to an overall amplitude $c$, the solution of this functional equation is
unique and given by $g(z) = c\ln z$. Choosing $c=1$ which a posteriori will
guarantee consistency with known special cases, we 
thus obtain the expression
\beq
\dsrt(t) =  k_B \ln\frac{p_\mr(t) \wrp}{p_\nr(t)\wrm} 
\ee
for the total entropy change induced by a forward transition $\rho$.
By separating the time-dependent part from the time-independent
one and by comparing with (\ref{eq:sss}), we can now identify both
(i) the missing piece in the total entropy change in one transition
 as
\beq
\Delta s_\rho(t) = - k_B \ln\frac{p_\nr(t)}{p_\mr(t)} 
\label{eq:stoch}
\ee
and  
(ii) a consistency relation between  the yet unknown
rates and the previously defined entropy change of medium and system
given by
\beq
\dsmr + \dsr= k_B \ln \frac\wrp\wrm .
\label{rates}
\ee 
Both identification make sense as we will show in the
next two subsections.

\subsection{Stochastic entropy}

Quite generally,
in a time-dependent ensemble specified by probabilities $p_n(t)$
stochastic entropy of the system  has been  defined as \cite{seif05a}
\beq
s(t) \equiv -\k \ln p_{n(t)}(t)
\label{eq:s}
\ee
along any individual trajectory $n(t)$ taken from the specified 
 ensemble. Hence, if a transition of type (\ref{reactm}) takes place at time $t$, the
full entropy change of the system $\Delta S\ss_\rho(t) $
consists of the change in stochastic entropy
(\ref{eq:stoch}) and that in intrinsic entropy
$\dsr$. Explicitly, one obtains  
\beq
\Delta S\ss_\rho(t)  \equiv   \Delta s_\rho (t)  + 
\dsr = -\k\ln\frac{\prp(t)}{\prm(t)} + \dsr.
\label{eq:ssys}
\ee 
Keeping the time argument  is
crucial since in a time-dependent ensemble the same transition leads to
a different contribution depending on when it takes place.

Finally, adding the concomitant change in entropy of the heat bath 
$\Delta S^{\rm med}_\rho=q_\rho /T$,
we obtain for the total entropy change associated
with this transition the expression (\ref{eq:sss}) which can also be written
as $
\Delta S^{\rm tot}(t) =  \Delta S^{\rm med}_\rho + \Delta S\ss(t) $. 
Note that in an equilibrium ensemble, where global detailed balance
applies, i.e., for 
$p_n(t) = p_n^{\rm eq}$ and 
$\prm^{\rm eq} \wrp=\prp^{\rm eq}\wrm$,  for
each jump the contribution to system entropy and medium entropy
exactly compensate each other so that the total entropy remains
strictly constant along any individual trajectory.

\subsection{Rates and local detailed balance}

The consistency relation derived above in (\ref{rates})
 between the ratio of the rates and
the sum of medium and system entropy change can be reformulated as
\beq
\frac{w_\rho^+}{w_\rho^-}  =\exp[-\beta \Delta F_\rho]= \exp[-\beta (\dfer + \Delta
\mu_\rho)], 
\label{eq:wrho}
\ee for the case of an enzymatic reaction and
as

\beq
\frac{w_\rho^+}{w_\rho^-}  = \exp[-\beta (\dfer + \Delta
\mu_\rho -  \wm)] ,
\label{eq:wrhom}
\ee
in the case of a motor protein where this transition involves external work, respectively.
Here, we have used (\ref{eq:sm}) and the first law in the form (\ref{eq:heatrelax}) and
(\ref{eq:qmot}), respectively. For molecular motors, the
additional exponential factors express the contribution of an applied force.
In both cases, all quantities depend on the concentrations $\{c_i\}$.

Both relations for the ratio of the rates of forward reaction to  backward reaction
are well known under the notion of ``local detailed balance''. In the present work, 
we have
shown that the rates have to obey this relation in order to get positive total
entropy production in a time-dependent ensemble.

\subsection{Dynamical formulation of the first law}

It is instructive to reformulate the two variants of the first law
discussed above for molecular motors
 in terms of these rates. For a closed system, prepared with
 non-equilibrium conditions,
 the heat released in this transition  becomes
\beq
q_\rho = T\left(\k   \ln \frac{w^+_\rho}{w^-_\rho} -\dsr\right) ,
\label{eq:qrho}
\ee
irrespective of whether external mechanical work is involved or not.
This expression shows that due to the presence of the intrinsic entropy
change $\dsr$ the heat dissipated in one transition cannot be infered
by just measuring the ratio of the rates.

For the alternative case of chemiostats with the explicit
refeeding and taking out of used and produced solutes
one gets
\beq
\bar q_\rho=   
T \left(\k \ln \frac{w^+_\rho}{w^-_\rho} - \dser \right) .
\label{eq:qrhocctwo}
\ee

For molecular motors often the case of a full cycle is discussed after
which the enzyme comes back to its initial internal state. Note that 
 for the scenario involving the chemiostats, the heat $\bar q$
dissipated along the cycle can then be expressed by  just the logarithm
of the ratio of the product of forward and backward rates along the cycle.
For the physically more realistic heat $q$, one has to correct for the entropy
change in the solution in order to determine  the heat from the
product of  the rates along a cycle.

\subsection{Fluctuation theorems}

So far, we have analyzed the changes in thermodynamic quantities caused
by an individual transition.
By summing up  the contributions from all transitions $\rho_j^{\sigma_j}$
happening during a time
interval $t_i<t<t_f$ and taking into account a possible change in stochastic entropy
$s(t)$ due to an explicit time-dependence
in $p_n(t)$ while the system stays in one state, one obtains the total entropy change along
a trajectory during this time interval as
\beq
 \Delta S^{\rm tot} =  \k \sum_j \ln
\frac{w_{{\rho_j}}^{\sigma_j}}{w_{{\rho_j}}^{-\sigma_j}}
+ s(t_f) - s(t_i) . 
\ee 
This quantity obeys a relation called the integral fluctuation theorem 
for entropy production  \cite{seif05a}
\beq
\langle \exp[-\Delta S^{\rm tot}/k_B]\rangle = 1 ,
\ee where the average $\langle ... \rangle$ is over many
trajectories taken from any well-defined initial ensemble
characterized by $p_n(t_i)$ and running for an arbitrary but fixed time
interval $t_f-t_i$. From this integral relation one
gets easily the second-law like statement on the
mean total
entropy production
\beq
\langle \Delta S^{\rm tot}
\rangle \geq 0
\label{eq:second}
.\ee
In the approach promoted in this paper, rather than ``deriving'' the latter relation
from the integral fluctuation theorem,
we have used it as an essential consistency requirement for
 (i) arguing that stochastic entropy is a crucial 
contribution to the entropy change occuring in an individual transition and
(ii) showing that rates obeying the  local detailed balance relations 
(\ref{eq:wrho},\ref{eq:wrhom})
 are required by thermodynamic
consistency.

For a non-equilibrium steady state where $p_n(t) = p_n$ is independent of
time, one  has
the detailed fluctuation theorem 
\beq p(- \Delta S^{\rm tot})/p( \Delta S^{\rm tot}) = \exp(- \Delta S^{\rm
  tot}/k_B)
\label{eq:dft}
\ee 
for the probability distribution $ p(\Delta S^{\rm tot})$ to 
observe a certain total entropy production
valid for any time interval in this non-equilibrium steady state \cite{seif05a}.

\section{Conclusions}
\label{sec6}

The present analysis is supposed to reveal more clearly
than previous work
both the structural coherence 
and some finer issues
of the stochastic thermodynamics  
of single enzymes and
molecular motors.
We first  summarize the main assumptions and consequences of this approach.

On the state level, internal energy, intrinsic entropy and
free energy follow from an underlying microscopic model if one assumes
 a time-scale separation between transitions within each state and
the slower and observable transitions between these states. It is not
necessary to assume  weak interactions between enzyme and solutes.
The first law with a concomitant identification of heat dissipated
in one transition then follows almost trivially. 
This form of the first law is in disagreement with formulations for 
motor proteins where chemiostats have been invoked as source of
``chemical work''. We have argued that their operation is somewhat artificial
as then is the corresponding identification of heat.

Entropy production on the level of an individual transition
still requires the notion of an ensemble. Enforcing the second law for any
time-dependent ensemble  then necessarily leads to  both stochastic
entropy as crucial contribution to the total entropy
change   and the local detailed balance condition for the rates. No
assumption, however, on the form of the individual forward and
backward rates are necessary. Hence, throughout the paper
there was no need to show ever  rates and
the master equation for the probabilities
$p_n(t)$ explicitly. Moreover, the present analysis shows that
the ``inverse'' problem is ill-posed: Given a master equation with transition
rates $\{w_{mn}\}$ (and, hence, a unique stationary state $\{p^s_n$\}), it is,
in general,
not possible to assign uniquely an internal energy level $E_n$, intrinsic
entropy $S_n$ and free energy $F_n$ to  a state. In particular, the choice
$E_n=-\kT\ln p^s_n$  suggested recently \cite{ge10} (like its obvious 
ramification $F_n=-\kT\ln p^s_n)$, while formally possible,
looks arbitrary and is, potentially, in conflict with an 
underlying specific microscopic model. If just transition rates
between states are given, only  a second law like statement about total entropy production
follows. Any further splitting up of the entropy production into  environment and system
let alone a definite form of the first law involving  internal energy and
exchanged heat is arbitrary
without further physical input.

The main difference of enzymes and motors compared 
to colloidal systems, which have no relevant hidden
 internal
degrees of freedom, is the crucial role the
intrinsic entropy of the states play. The latter shows up if the first law
is expressed dynamically through the rates. It prevents a direct inference
of the dissipated heat from a measurement of the ratio of the rates
even on the level of a  complete cycle. In the total
entropy production, on the other hand, the intrinsic entropy does not appear
explicitly. Therefore, the fluctuation theorems hold true unmodified.
A consequence, of course, is that these theorems cannot be used 
to ``disentangle'' the dissipated heat from the entropy change occuring in the
solution.

Throughout the paper, we have assumed stationary non-equilibrium conditions
which implies that the rates are time-independent.
In fact, the results for the individual transitions hold true even if the
rates become time-dependent either since the concentrations of the solutes
are externally modulated (or, in a finite system, depleted due to the action of the
enzymes) or since the forces applied to motor proteins 
are time-dependent. However, under such time-dependent external conditions
the thermodynamic state variables $E_n,S_n$ and $F_n$ can become time-dependent as well.
In consequence, both the quantities
appearing in the first law and entropy production can pick up contributions
even while the enzyme remains in the same state.

Finally, with the conceptual basis thus solidified, 
these thermodynamic notions should now be applied to  data
from single enzyme experiments. A very
promising molecule seems to the F1-ATPase for which the first 
experimental studies using
such concepts have just appeared \cite{toya10,haya10}.

\section {Appendix: Non-equilibrium ensemble thermodynamics}

The main intention of the approach discussed in this paper has been
the identification of thermodynamic quantities not on the ensemble level
but for a single enzyme along its fluctuating
trajectory taken from a well-defined ensemble. For completeness and future
reference, we briefly present the
consequences of our thermodynamically consistent approach for  time-dependent
averages. 

Quite generally, once for each state $n$
of a system an internal energy $E_n$, an intrinsic entropy $S_n$,
and a free energy $F_n$ are identified from a more microscopic model, 
the ensemble average of the internal energy
becomes
\beq
{\cal E}(t) \equiv \langle E(t)\rangle = \sum_n p_n(t) E_n .
\label{eq:ce}
\ee 
The appropriate ensemble averaged entropy of the system is given by
 \beq
{\cal S}(t)  \equiv \sum_n p_n(t)[ S_n + s_n(t)],
\label{eq:cs}
\ee
with
$
s_n(t) \equiv -\k \ln p_n(t)$.
A ``dynamical free energy'' on the ensemble level then follows from the usual
thermodynamic relation as
\beq
{\cal F} (t) \equiv  {\cal E}(t) - T {\cal S}(t) =  \sum_n p_n(t)[ F_n  -T s_n(t) ].
\label{eq:fs}
\ee 
Note that both quantities, ${\cal S}(t)$ and ${\cal F} (t)$, cannot be
obtained by simply averaging over the state variable $S_n$ or $F_n$
as it is possible for
the internal energy. The
expression $s_n(t)$ arising from 
stochastic entropy in the square brackets in (\ref{eq:cs}) and (\ref{eq:fs}) 
is rather ensemble dependent and thus not a genuine state variable.

For a consistency check of these ensemble expressions, consider an equilibrium ensemble.
Then the probability to find the system in any microstate $\xi$
is given by
\beq
\peqx = \exp[-\beta(V^{\rm tot}(\xi)-\Feq)] .
\ee
with the equilibrium free energy 
\beq
\Feq\equiv -\kT \ln  \sum_\xi\exp[-\beta V^{\rm tot}(\xi)] .
\label{eq:F}
\ee 
for the total system consisting of enzyme including the surrounding 
solution.
Likewise, one has for the ensemble internal energy
\beq
\Ee = \sum_{\x} \peqx V^{\rm tot}(\xi)
\ee
and the ensemble system entropy
\beq
\Se = -\k \sum_{\x} \peqx\ln \peqx = (\Ee-\Feq)/T  .
\ee
It is easily checked that the time-dependent quantities (\ref{eq:ce}-\ref{eq:fs})
agree with these 
equilibrium ensemble quantities, if $p_n(t)$ in (\ref{eq:ce}-\ref{eq:fs})
becomes the
equilibrium probability
\beq
\peqn\equiv \exp[-\beta (\F_n-\Feq)] .
\ee with $F_n$ as previously defined in (\ref{eq:fnn}).
 
For the rate of change of the internal energy of the system, i.e., here the enzyme or motor
plus the surrounding solution,  one obtains
\beq
\dot {\cal E}(t)  = \sum_\rho (\prm(t)\wrp-\prp(t)
\wrm)\der .
\label{eq:ced}
\ee 
This expression enters the rate formulation of the first law  on the
ensemble level as given by
\beq
\dot W^{\rm mech}(t) = \dot Q(t) +  \dot {\cal E}(t)
\ee
with
\beq
\dot W^{\rm mech}(t) = \sum_\rho (\prm(t)\wrp-\prp(t)
\wrm) \wm
\ee
as the ensemble averaged work rate.
The heat dissipation rate on the ensemble level thus becomes
\beq
\dot Q(t) = \sum_\rho (\prm(t)\wrp-\prp(t) 
\wrm)q_\rho .\ee 
Likewise, for the full entropy change of the system, one obtains
 \beq
\dot {\cal S}(t) =  
\sum_\rho (\prm(t)\wrp-\prp(t)
\wrm)\dsr^{\rm sys} (t) ,
\ee
with $\dsr^{\rm sys} (t)$ from (\ref{eq:ssys})
and for the corresponding change in free energy 
\begin{eqnarray}
\dot {\cal F} (t) &=& \dot {\cal E}(t) - T \dot{\cal S}(t) \\&=& 
\sum_\rho (\prm(t)\wrp-\prp(t)
\wrm) \times\\
&~&~~~~~~ \times (\der - T \dsr^{\rm sys}(t)) .
\label{eq:fsd}
\end{eqnarray}
Note that if no external work is applied, the thus defined free energy dissipation
rate
is exactly given by (-$T$) times  the total entropy production rate (\ref{eq:ssss}).





\end{document}